\documentclass{article}
\usepackage{spconf,amsmath,graphicx,url}

\title{The DIRHA-English corpus and related tasks for distant-speech recognition in domestic environments}
\name{Mirco Ravanelli, Luca Cristoforetti, Roberto Gretter,  Marco Pellin, Alessandro Sosi, Maurizio Omologo}
\address{Fondazione Bruno Kessler (FBK), 38123 Povo, Trento, Italy \\ 
\small \tt (mravanelli,cristofo,gretter,pellin,alesosi,omologo)@fbk.eu}

\begin{document}
%
\maketitle
\begin{abstract}

This paper introduces the contents and the possible usage of the DIRHA-ENGLISH multi-microphone corpus, recently realized under the EC DIRHA project. The reference scenario is a domestic environment equipped with a large number of microphones and microphone arrays distributed in space.

The corpus is composed of both real and simulated material, and it includes 12 US and 12 UK English native speakers. 
Each speaker uttered different sets of phonetically-rich sentences, newspaper articles, conversational speech, keywords, and commands. 
From this material, a large set of 1-minute sequences was generated, which also includes typical domestic background noise as well as inter/intra-room reverberation effects.
Dev and test sets were derived, which represent a very precious material for different studies on multi-microphone speech processing and distant-speech recognition. 
Various tasks and corresponding Kaldi recipes have already been developed. 

The paper reports a first set of baseline results obtained using different techniques, including Deep Neural Networks (DNN), aligned with the state-of-the-art at international level. 
\end{abstract}
\section{Introduction}
\label{sec:intro}
During the last decade, much research has been devoted to improve Automatic Speech Recognition (ASR) performance \cite{lideng}. As a result, ASR has recently been applied in several fields, such as web-search, car control, automated voice answering, radiological reporting and it is currently  used by millions of users worldwide. 
Nevertheless, most state-of-the-art systems are still based on close-talking solutions, forcing the user to speak very close to a microphone-equipped device. 
Although this approach usually leads to better performance, it is easy to predict that, in the future, users will prefer to relax the constraint of handling or wearing any device to access speech recognition services. There are indeed various real-life situations where a distant-talking (far-field)\footnote{In the following, the same concept is referred to as ``distant-speech".} interaction is more natural, convenient and attractive \cite{dasr}.
In particular, amongst all the possible applications, an emerging field is speech-based domestic control, where users might prefer to freely interact with their home appliances without wearing or even handling any microphone-equipped device. 
This scenario was addressed under the EC DIRHA (Distant-speech Interaction for Robust Home Applications) project\footnote{The research presented here has been partially funded by the
European Unionâs 7th Framework Programme (FP7/2007-2013) under grant agreement no. 288121 DIRHA (for more details, please see http:// dirha.fbk.eu).}, which had the ultimate goal of developing voice-enabled automated home services based on Distant-Speech Recognition (DSR) in different languages.

Despite the growing interest towards DSR, current technologies still exhibit a significant lack of robustness and flexibility, since the adverse acoustic conditions originated by non-stationary noises and acoustic reverberation make speech recognition significantly more challenging \cite{adverse}. 
Although considerable progresses were made at multi-microphone front-end processing level in order to feed ASR with an enhanced speech input \cite{BrandWard,beam,nadeu,bss,derev}, the performance loss observed from close-talking to distant-speech remains quite critical, even when the most advanced DNN-based backend frameworks are adopted \cite{pawel2,hain,dnn_rev,dnn_rev2,dnn3,rav_in15}.

To further progress, a crucial step regards the selection of data suitable to train and test the various speech processing, enhancement, and recognition algorithms. 
Collecting and transcribing sufficiently large data sets to cover any possible application scenario is a prohibitive, time-consuming and expensive task. In a domestic context, in particular, due to the large variabilities that can be introduced when deploying such systems in different houses, this issue becomes even more challenging than in any other traditional ASR application. 
In this context, an ideal system should be flexible enough in terms of microphone distribution in space, and in terms of other possible profiling actions. Moreover, it must be able to provide a satisfactory behaviour immediately after its installation, and to improve performance thanks to its capability to learn from the environment and from the users.


In order to develop such solutions, the availability of high-quality and realistic, multi-microphone corpora represents one of the fundamental steps towards reducing the performance gap between close-talking and distant-speech interaction.
Along this direction,
strong efforts have been spent recently by the international scientific community, through the development of corpora and challenges, such as REVERB \cite{revch}, CHIME \cite{chime,chime3} and ASpIRE.
Nevertheless, we feel that other complementary corpora and tasks are necessary to the research community in order to further boost technological advances in this field, for instance providing a large number of ``observations" of the same acoustic scene.

The DIRHA-ENGLISH corpus complements the set of corpora previously collected under the DIRHA project in other four languages (i.e., Austrian German, Greek, Italian, Portuguese) \cite{lrec}. It gives the chance of working on English, the most commonly used language inside the ASR research community, with a very large number of microphone channels, a multi-room setting, and the use of microphone arrays having different characteristics. Half of the material is based on simulations, and half is based on real recordings, which allows one to assess recognition performance in real-world conditions.
It is also worth mentioning that some portions of the  corpus will be made publicly available, with free access, in the short term (as done with other data produced by the DIRHA consortium).

The purpose of this paper is to introduce the DIRHA-ENGLISH corpus as well as to provide some baseline results on phonetically-rich sentences, which were obtained using the Kaldi framework \cite{kaldi}.  
The resulting TIMIT-like phone recognition task can be seen as complementary to the WSJ-like and conversational speech tasks also available for a next distribution. 




The remainder of the paper is organized as follows. Section \ref{sec:dirha} provides a brief description of the DIRHA project, while Section \ref{sec:dirha_english} focuses the contents and characteristics of the DIRHA-ENGLISH corpus. Section \ref{sec:exp} gives a description of the experimental tasks so far defined, and of the corresponding baseline results. Section \ref{sec:concl} draws some conclusions.

\section{The DIRHA project}
\label{sec:dirha}
The EC DIRHA project, which started in January 2012 and lasted three years, had the goal of addressing acoustic scene analysis and distant-speech interaction in a home environment. 
In the following, some information are reported about project goals, tasks, and corpora. 

\subsection{Goals and tasks}
The application scenario targeted by the project is characterized by a quite flexible voice interactive system to talk with in any room, and from any position in space. Exploiting a microphone network distributed in the different rooms of the apartment, the DIRHA system reacts properly when a command is given by the user. The system is always-listening, waiting for a specific keyword to ``capture" in order to begin a dialogue. The dialogue that is triggered in this way, gives the end-user a possible access to devices and services, e.g., open/close doors and windows, switch on/off lights, control the temperature, play music, etc.
Barge-in (to interact while music/speech prompts are played), speaker verification, concurrent dialogue management (to support simultaneous dialogues with different users) are some advanced features characterizing the system.
Finally, a very important aspect to mention is the need to limit the rate of false alarms, due to possible misinterpretation of normal conversations or of other sounds captured by the microphones, which do not carry any relevant message to the system.

Starting from these targeted functionalities, several experimental tasks were defined concerning the combination between front-end processing algorithms (e.g., of speaker localization, acoustic echo cancellation, speech enhancement, etc.) and an ASR backend, in each language.
Most of these tasks were referred to voice interaction in the ITEA apartment\footnote{We would like to thank ITEA S.p.A (Istituto Trentino per l'Edilizia Abitativa) for making available the apartment used for this research.},  situated in Trento (Italy), which was the main site for acoustic and speech data collection.






\subsection{DIRHA corpora} 
\label{...}

The DIRHA corpora were designed in order to provide multi-microphone data sets that can be used to investigate a wide range of tasks as those mentioned above.

Some data sets were based on simulations realized applying a contamination method \cite{matassoni,cont2,cont3} that combines clean-speech signals, estimated Impulse Responses (IRs), and real multichannel background noise sequences, as described in \cite{lrec}.
Other corpora were recorded under real-world conditions.

Besides the DIRHA-ENGLISH corpus described in the next section, other corpora developed in the project are:
\begin{itemize}
\item The DIRHA Sim corpus described in \cite{lrec} (30 speakers x 4 languages), which consists of 1-minute multi-channel sequences including different acoustic events and speech utterances; 
\item A Wizard-of-OZ data set proposed in \cite{hscma} to evaluate the performance of speech activity detection and speaker localization components;
\item The DIRHA AEC corpus \cite{zwyssig2015_dirha}, which includes data specifically created for studies on Acoustic Echo Cancellation (AEC), to suppress known interferences diffused in the environment (e.g., played music);
\item The DIRHA-GRID corpus \cite{dirha_grid}, a multi-microphone multi-room simulated data set that derives from contaminating the GRID corpus \cite{grid} of short commands in the English language.
\end{itemize}

\section{The DIRHA-ENGLISH corpus}
\label{sec:dirha_english}

As done for the other four languages, also the DIRHA-ENGLISH corpus consists of a real and a simulated data set, the latter one deriving from contamination of a clean speech data set described next.

\subsection{Clean speech material}
\label{sec:clean}
The clean speech data set was realized in a recording studio of FBK, 
using professional equipment (e.g., a Neumann TLM 103 microphone) to obtain high-quality 96 kHz - 24 bit material. 

12 UK and 12 US speakers were recorded (6 males and 6 females, for each language). 
For each of them, the corresponding recorded material includes: 
\begin{itemize}
\item 15 read commands;
\item 15 spontaneous commands;
\item 13 keywords; 
\item 48 phonetically-rich sentences (from the Harvard corpus); 
\item 66 or 67 sentences from WSJ-5k; 
\item 66 or 67 sentences from WSJ-20k; 
\item about 10 minutes of conversational speech (e.g., the subject was asked to talk about a movie recently seen).
\end{itemize}

The total time is about 11 hours. All the utterances were manually annotated by an expert.
For the phonetically-rich sentences, an automatic phone segmentation procedure was applied as done in \cite{segmentazione}. An expert then checked manually the resulting phone transcriptions and time-aligned boundaries to confirm their reliability.

For both US and UK English, 6 speakers were assigned to the development set, while the other 6 speakers were assigned to the test set. These assignments were done in order to distribute WSJ sentences as in the original task \cite{wsj_design}. The data set contents are compliant with TIMIT specifications (e.g., file format). 

\begin{figure}[t!]
\centering
\includegraphics[width=0.49\textwidth]{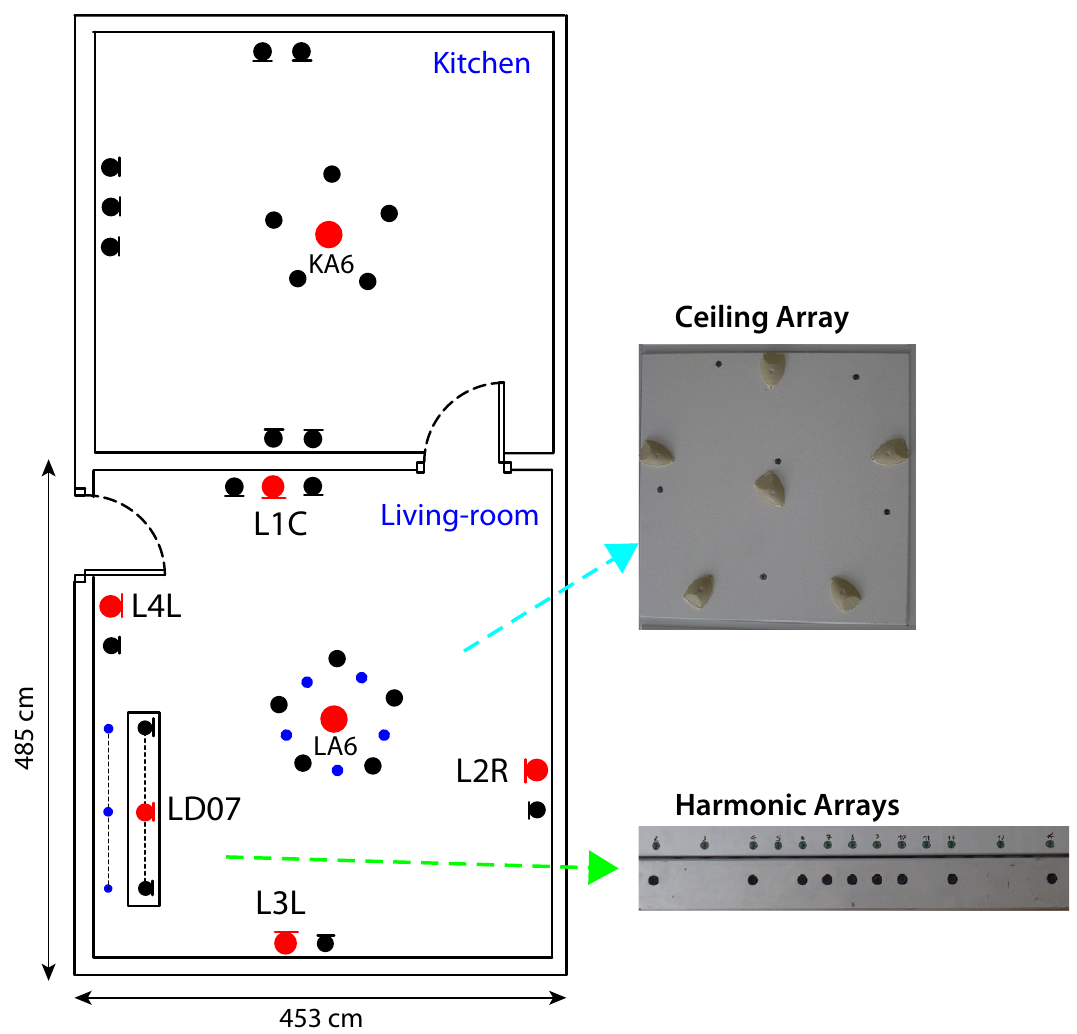}
\caption{An outline of the microphone set-up adopted for the DIRHA-ENGLISH corpus. Blue small dots represent digital MEMS microphones, red ones refers to the channels considered for the following experimental activity, while black ones represent the other available microphones. The right pictures show the ceiling array and the two linear harmonic arrays installed in the living-room.}
\label{fig:dirhaflat}
\end{figure}


\subsection{The microphone network} 
\label{sec:itea}
The ITEA apartment   is the reference home environment that was available during the project for data collection as well as for the development of 
prototypes and showcases. The flat comprises five rooms which are equipped with a network of several microphones.
Most of them are high-quality omnidirectional microphones (Shure MX391/O), connected to multichannel clocked pre-amp and A/D boards (RME Octamic II), which allowed a perfectly synchronous sampling at 48 kHz, with 24 bit resolution.
The bathroom and two other rooms were equipped with a limited number of microphone pairs and triplets (i.e., overall 12 microphones), while the living-room and the kitchen comprise the largest concentration of sensors and devices. As shown in Figure \ref{fig:dirhaflat}, the living-room includes three microphone pairs, a microphone triplet, two 6-microphone ceiling arrays (one consisting of MEMS digital microphones), two harmonic arrays (consisting of 15 electret microphones and 15 MEMS digital microphones, respectively). 
More details about this facility can be found in \cite{lrec}.



Concerning this microphone network, a strong effort was devoted to characterize the environment at acoustic level, through different campaigns of IR estimation, leading to more than 10.000 IRs that describe sound propagation from different positions in space (and different possible orientations of the sound source) to any of the available microphones.
The method adopted to estimate an IR consists in diffusing  
a known Exponential Sine Sweep (ESS) signal in the target environment, and recording it by the available microphones \cite{farina}.
The accuracy of the resulting IR estimation has a remarkable impact on the speech recognition performance, as shown in \cite{ravanelli}.
For more details on the creation of the ITEA IR database, please refer to \cite{ravanelli,lrec}. Note that the microphone network considered for the DIRHA-ENGLISH data set (shown in Fig.\ref{fig:dirhaflat}) is limited to the living-room and to the kitchen of the ITEA apartment, but also considers harmonic arrays and MEMS microphones which were unavailable in the other DIRHA corpora.

\subsection{Simulated data set}

The DIRHA-ENGLISH simulated data sets derive from the clean speech described in Section \ref{sec:clean}, and from the application of the contamination method discussed in \cite{matassoni,ravanelli}.

The resulting corpus consists of a large number of 1-minute sequences, each including a variable number of sentences uttered in the living-room with different noisy conditions. 
Four types of sequences have been created, corresponding to the respective following tasks: 1) Phonetically-rich sentences\footnote{While noisy conditions are quite challenging for the  WSJ and conversational parts of the DIRHA-English corpus, the phonetically-rich sequences are characterized by more favorable conditions in order to make this material more suitable for studies on reverberation effects only.}; 2) WSJ 5-k utterances; 3) WSJ 20-k utterances; 4) Conversational speech (also including keywords and commands).

For each sequence, 62 microphone channels are available, as outlined in Section \ref{sec:itea}.

\subsection{Real data set}

For what concerns real recordings, each subject was positioned in the living-room and read the material from a tablet, standing still or sitting on a chair, in a given position. After each set, she/he was asked to move to a different position and take a different orientation.
For each speaker, the recorded material corresponds to the same list of contents reported in Section \ref{sec:clean} for the clean speech data set.

Note also that all the channels recorded through MEMS digital microphones were time-aligned with the others during a post-processing step (since using the same clock for all the platforms was not feasible due to different settings and sampling frequency in the case of MEMS devices) .

Once collected the whole real material, 1-minute sequences were derived from it in order to ensure a coherence in terms of sequence between simulation and real data sets.

\section{Experiments and Results}
\label{sec:exp}
This section describes the proposed task and the related baseline experiments concerning the US phonetically-rich portion of the DIRHA-English corpus. 

\subsection{ASR framework}
\label{sec:asr}

\subsubsection{Training and testing corpora}
\label{sec:tr_te_corpora}

In this work, the training phase is accomplished with the train part of the TIMIT corpus \cite{timit}. 
For the DSR experiments, the original TIMIT dataset is reverberated using three impulse responses measured in the living-room. Moreover, some multi-channel noisy background sequences are added to the reverberated signals, in order to better match real-world conditions. Both the impulse responses and the noisy sequences are different from those used to generate the DIRHA-ENGLISH simulated data set.

The test phase is conducted using the real and the simulated phonetically-rich sentences of the DIRHA-English data set. 
In both cases, for each 1-minute sequence an oracle voice activity detector (VAD) is applied in the next experiments\footnote{Alternative tasks (not presented here) have been defined as well with the same material, in order to investigate VAD and ASR components together.}, in order to avoid any possible bias due to inconsistent sentence boundaries. 
A down-sampling of the speech sequences from 48 to 16 kHz is finally performed. 

\subsubsection{Feature extraction}
\label{sec:features}
A standard feature extraction based on MFCCs is applied to the speech sentences. In particular, the signal is blocked into frames of 25 ms with 10 ms overlapping and, for each frame, 13 MFCCs are extracted. The resulting features, together with their first and second order derivatives, are then arranged into a single observation vector of 39 components. 

\subsubsection{Acoustic model training}
\label{sec:am_training}
In the following experiments, three different acoustic models of increasing complexity are considered. The procedure adopted for training such models is the same as that used for the original s5 TIMIT Kaldi recipe  \cite{kaldi}. The first baseline (\textit{mono}), refers to a simple system characterized by 48 context-independent phones of the English language, each modeled by a three state left-to-right HMM (overall using 1000 gaussians). 
The second baseline (\textit{tri}) is based on a context-dependent phone modeling and on speaker adaptive training (SAT). 
Overall, 2.5k tied states with 15k gaussians are employed. Finally, the DNN baseline (\textit{DNN}), trained with the Karel's recipe \cite{karel}, 
is composed of 6 hidden layers of 1024 neurons,  with a context window of 11 consecutive frames (5 before and 5 after the current frame) and an initial learning rate of 0.008.


\subsubsection{Proposed task and evaluation}
\label{sec:proposed_task}
The original Kaldi recipe is based on a  bigram language model estimated from the phone transcriptions available in the training set. Conversely, we propose the adoption of a pure phone-loop (i.e., zero-gram based) task, in order to avoid any non-linear influence and artifacts possibly originated by a language model. Our past experience \cite{ravanelli,rav_in14,rav_in15} indeed suggests that, even though the use of language models is certainly helpful in increasing the recognition performance, the adoption of a simple phone-loop task is more suitable for experiments purely focusing on the acoustic information.

\begin{figure}[t]
\centering
\includegraphics[height=8cm, width=7cm]{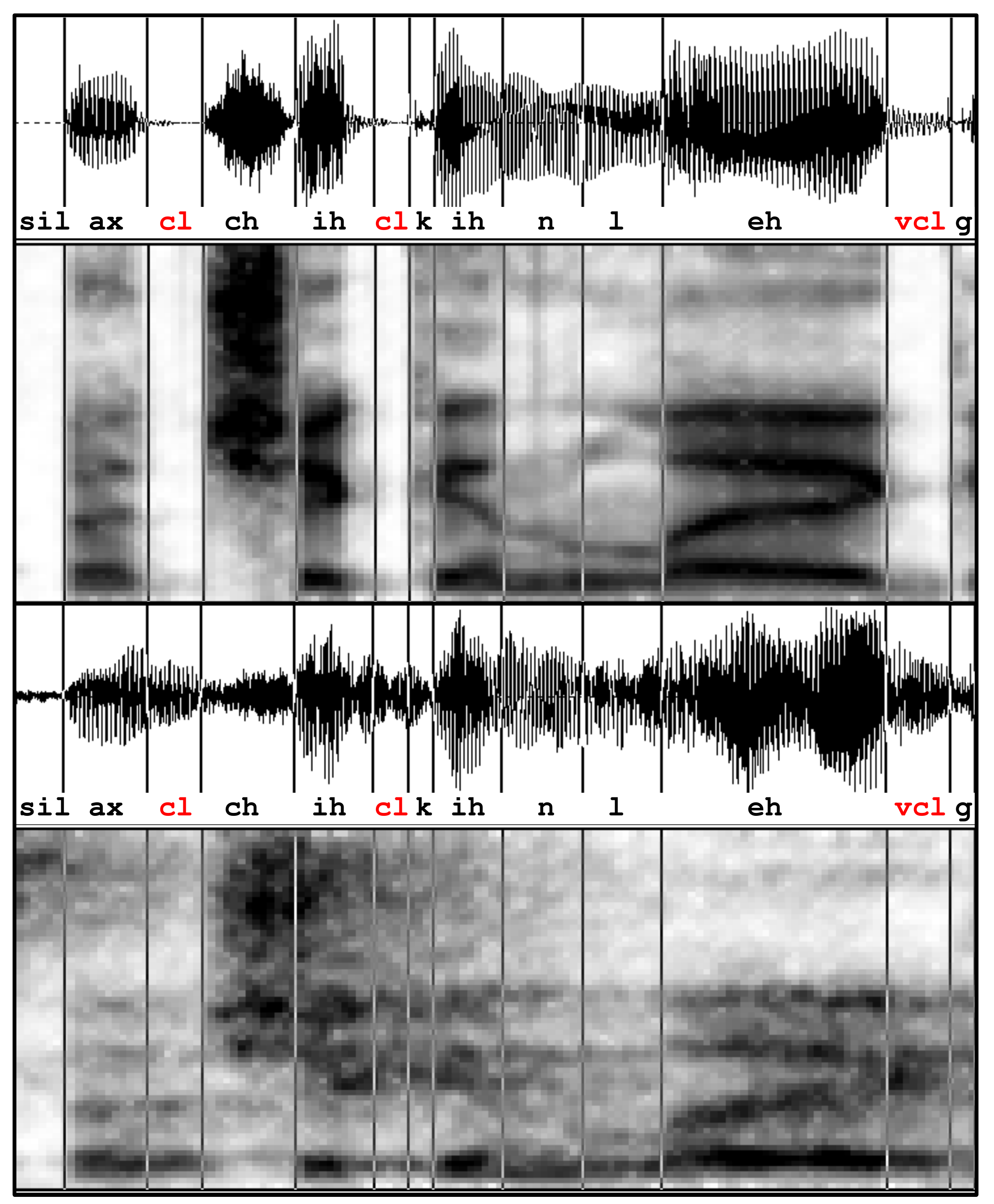}
\caption{The phrase ``\textit{a chicken leg}" uttered in close and distant-talking scenarios, respectively. The closures (in red) are dimmed by the reverberation tail in the distant speech.}
\label{fig:comparison}
\end{figure}

Another difference with the original Kaldi recipe regards the evaluation of silences and closures. In the evaluation phase, the standard Kaldi recipe (based on \textit{Sclite}) maps the original 48 English phones into a reduced set of 39 units, as originally done in \cite{kflee}. In particular, the six closures (\textit{bcl}, \textit{dcl}, \textit{gcl}, \textit{kcl}, \textit{pcl}, \textit{tcl}) are mapped as ``optional silences" and possible deletions of such units are not scored as errors. These phones would be likely considered as correct,  since deletions of short closures occur very frequently. 
We believe that the latter aspect might introduce a bias in the evaluation metrics, especially for DSR tasks, where the reverberation tail makes the recognition of the closures nearly infeasible, as highlighted in Figure \ref{fig:comparison}. For this reason, we propose to simply filter out all the silences and closures from both the reference and the hypothesized  phone sequences. This leads to a performance reduction, since all the favorable optional silences added in the original recipe are avoided. However, a more coherent estimation of the recognition rates concerning phones as occlusive and vowels is reached.



\subsection{Baseline results}
\label{sec:baselines}
This section provides some baseline results\footnote{Part of the experiments are conducted with a Tesla K40 donated by the NVIDIA Corporation.}, which might be useful in the future to other researchers for reference purposes.
In the following sections, results based on close-talking and distant-speech input are presented.

\subsubsection{Close-talking performance}
\label{sec:ct_res}
Table \ref{tab:ct_res1} reports the performance obtained by decoding the clean sentences recorded in the FBK recording studio with either a phone bigram language model or a simple phone-loop. Results are provided using both the standard Kaldi s5 and the proposed recipe, in order to highlight all the discrepancies in performance that can be observed in these different experimental settings.

\begin{table}[h!]
\centering
\small
    \begin{tabular}{ | l | c | c | c | c | }
    \hline 
      Recipe & LM Type & Mono  & Tri & DNN   \\ \hline
      Standard Kaldi s5 & Bigram LM & 36.4  & 23.2 & 20.1  \\ \hline
      Standard Kaldi s5 & Phone-loop & 39.4  & 26.3 & 22.4   \\ \hline
      Proposed Evaluation & Bigram LM & 42.7  & 28.6 & 24.6   \\ \hline
      Proposed Evaluation & Phone-loop & 46.7  & 32.5 & \textbf{27.5}   \\ \hline

 \end{tabular}
\caption{Phone Error Rate (PER\%) performance obtained applying different Kaldi recipes to the phonetically-rich sentence (dev-set) acquired in the FBK recording studio.}
\label{tab:ct_res1}
\end{table}

As expected, these results highlight that the system performance is significantly improved when passing from a simple monophone-based model to a more competitive DNN baseline.
Moreover, as outlined in Sec.\ref{sec:proposed_task}, applying the original Kaldi evaluation provides a mismatch of about 20\% in relative error reduction, which does not correspond to any real system improvement.  
Next experiments will be based on the pure phone-loop grammar scored with the proposed evaluation method.




\subsubsection{Single distant-microphone performance}
In this section, the results obtained with a single distant microphone are discussed. Table \ref{tab:dist_res1} shows the performance achieved with some of the microphones highlighted in  Fig.\ref{fig:dirhaflat}.

\begin{table}[h!]
\centering
\small
    \begin{tabular}{ | c | c | c | c | c | c | c | }
    \hline
     & \multicolumn{3}{| c |}{Simulated Data} & \multicolumn{3}{| c |}{Real Data} \\ \cline{2-7}
      Mic. ID & Mono  & Tri & DNN & Mono  & Tri & DNN   \\ \hline
      LA6 & 68.8  & 57.7 & \textbf{51.6} & 70.5  & 60.9 & \textbf{55.1}   \\ \hline
      L1C & 67.4  & 58.5 & 52.4 & 70.3  & 61.7 & 55.6   \\ \hline
      LD07 & 67.5 & 58.1 & 53.2 & 71.5  & 62.6 & 57.3   \\ \hline
      KA6 & 76.7  & 67.3 & 64.0 & 80.5  & 73.6 & 70.5   \\ \hline
      
 \end{tabular}
\caption{PER(\%) performance obtained with single distant microphones for both the simulated and real dataset of phonetically-rich sentences.}
\label{tab:dist_res1}
\end{table}

The results clearly highlights that in the case of distant-speech input the ASR performance is dramatically reduced, if compared to a close-talking case. 
As already observed with close-talking results, the use of a DNN significantly outperforms the other acoustic modeling approaches. This is consistent for all the considered channels, with both simulated and real data sets.  Actually, the performance on real data is slightly worse than that achieved on simulated data, due to a lower SNR characterizing the real recording sessions. 

It is also worth noting that almost all the channels provide a similar performance and a comparable trend over the considered acoustic models. Only the kitchen microphone (KA6) corresponds to a more challenging situation, since all the utterances of both real and simulated data sets were pronounced in the living-room.



\subsubsection{Delay-and-sum beamforming performance}
This section reports the results obtained with a standard delay-and-sum beamforming \cite{BrandWard} applied to both the ceiling and the harmonic arrays of the living-room. 

\begin{table}[h!]
\centering
\small
    \begin{tabular}{ | l | c | c | c | c | c | c | }
    \hline
     & \multicolumn{3}{| c |}{Simulated Data} & \multicolumn{3}{| c |}{Real Data} \\ \cline{2-7}
      Array ID & Mono  & Tri & DNN & Mono  & Tri & DNN   \\ \hline
      Ceiling arr. & 66.2  & 55.9 & \textbf{50.4} & 65.9  & 55.9 & \textbf{50.6}   \\ \hline
      Harmonic arr. & 66.2  & 56.0 & 51.8 & 66.2  & 56.2 & 51.5   \\ \hline
 \end{tabular}
\caption{PER(\%) performance obtained with a delay-and-sum beamforming applied to both the ceiling and the linear harmonic array.}
\label{tab:beam_res}
\end{table}

Table \ref{tab:beam_res} shows that beamforming is helpful in improving the system performance. For instance, in the case of real data one passes from a PER of 55.1$\%$, with the single microphone, to a PER of 50.6$\%$, when delay-and-sum beamforming is applied to the ceiling array signals. 
 
Even though the ceiling array is composed of six microphones only, it ensures a slightly better performance when compared with a less compact 13 element harmonic array. This result might be due  both to a better position of the ceiling array, which often ensures the presence of a direct path stronger than reflections,  and to adoption of  higher quality microphones. 
The performance improvement introduced by delay-and-sum beamforming is higher with real data, confirming that
spatial filtering techniques are particularly helpful when the acoustic conditions are less stationary and predictable.

\subsubsection{Microphone selection performance}
The DIRHA-English corpus can also be used for microphone selection experiments. It would be thus of interest to provide some lower and upper bound performance for a microphone selection technique applied to this data set. Table \ref{tab:mic_sel} compares the results achieved with random and with oracle selections of the microphone, for each phonetically-rich sentence. For this selection, we considered the six microphones of the living-room, which are depicted as red dots in Figure \ref{fig:dirhaflat}.

\begin{table}[h!]
\centering
\small
    \begin{tabular}{ | c | c | c | c | c | c | c | }
    \hline
     & \multicolumn{3}{| c |}{Simulated Data} & \multicolumn{3}{| c |}{Real Data} \\ \cline{2-7}
      Mic. Sel. & Mono  & Tri & DNN & Mono  & Tri & DNN   \\ \hline
      Random & 67.6  & 57.7 & 52.4 & 70.3  & 61.0 & 55.4   \\ \hline
      Oracle & 56.6  & 47.1 & \textbf{42.0} & 60.3  & 49.6 & \textbf{44.0}   \\ \hline
 \end{tabular}
\caption{PER(\%) performance obtained with a random and an oracle microphone selection.}
\label{tab:mic_sel}
\end{table}

Results show that a proper microphone selection is crucial for improving the performance of a DSR system. The gap between the upper bound limit, based on an oracle channel selection and the lower bound limit, based on a random selection of the microphone, is particularly large. This confirms the importance of suitable microphone selection criteria. 
A proper channel selection has a great potential even when compared with a microphone combination based on delay-and-sum beamforming. For instance, a PER of 44.0$\%$ is obtained with an oracle channel selection against a PER of 50.6\% achieved with the ceiling array. 

\section{Conclusions and future work}
\label{sec:concl}

This paper described the DIRHA-ENGLISH multi-microphone corpus and the first baseline results concerning the use of the phonetically-rich sentence data sets.
Overall, the experimental results show the expected trend of performance, quite well aligned to past works in this field.

In research studies on DSR, there are many advantages
in using phonetically-rich material with such a large number of microphone channels. For instance, there is the possibility of better focusing on the impact on performance of some front-end processing techniques for what concerns specific phone categories.

The corpus, also includes WSJ and conversational speech data 
sets that can be object of public distribution\footnote{The distribution of WSJ data set is under discussion with LDC.}
and of a possible use in next challenges regarding DSR.
The latter data sets can be very helpful to investigate other key topics as, for instance, multi-microphone hypothesis combination based on confusion networks, multiple lattices, and rescoring. 
Forthcoming works include the development of baselines and related recipes for MEMS microphones, for  WSJ and conversational sequences, as well as for the UK English language. 

\section{Corpus release}
\label{sec:release}
Some 1-minute sequences can be found at \url{http://dirha.fbk.eu/DIRHA_English}.
The access to data that were used in this work, and to related documents, will be possible soon through a FBK server, with modalities that will be reported under http://dirha.fbk.eu. 
In the future, other data sets will be made publicly available, together with corresponding documentation and recipes, and with instructions to allow comparison of systems and maximize scientific insights. 


\bibliographystyle{IEEEbib}
\bibliography{refs}

\end{document}